# Nanoscale NMR Spectroscopy using Self-Calibrating Nanodiamond Quantum Sensors

Jeffrey Holzgrafe[1,2,*,†], Qiushi Gu[1,*], Jan Beitner[1], Dhiren Kara[1], Helena S. Knowles[1,3], and Mete Atatüre[1†]

[1] *Cavendish Laboratory, University of Cambridge, JJ Thomson Avenue, Cambridge CB3 0HE, UK*

[2] *John A. Paulson School of Engineering and Applied Sciences, Harvard University 29 Oxford Street, Cambridge MA 02138, USA*

[3] *Department of Physics, Harvard University, 17 Oxford Street, Cambridge MA 02138, USA*

*: *Equal contribution*

[†] Corresponding authors: J.H. (jholzgrafe@g.harvard.edu), M.A. (ma424@cam.ac.uk).

**Abstract**

Conventional nuclear magnetic resonance (NMR) spectroscopy relies on acquiring signal from a macroscopic ensemble of molecules to gain information about molecular structure and dynamics. Transferring this technique to nanoscale sample sizes would enable molecular analysis without the effects of averaging over spatial and temporal inhomogeneities and without the need for macroscopic volumes of analyte, both inherent to large ensemble measurements. Nanoscale NMR based on nitrogen vacancy (NV) centers inside bulk diamond chips achieves single nuclear spin sensitivity and the resolution required to determine chemical structure, but their detection volume is limited to a few nanometers above the diamond surface for the most sensitive devices. This precludes them from use for nuclear spin sensing with nanoscale resolution inside thicker structures, such as cells. Here, we demonstrate the detection of NMR signals from multiple nuclear species in a $\sim(19\ \text{nm})^3$ volume using versatile NV-NMR devices inside nanodiamonds that have a typical ~30 nm diameter. The devices detect a signal generated by a small number of analyte molecules on the order of ~1000. To use these devices *in situ*, the detected signal must be corrected for the unknown geometry of each nanodiamond device. We show that such a calibration could be performed by exploiting the signal from a thin layer of nuclei on the diamond surface. These results, combined with nanodiamonds' low toxicity and amenability to surface functionalization, indicate that nanodiamond NV-NMR devices could become a useful tool for nanoscale NMR-based sensing inside living cells.

**Introduction**

Nuclear magnetic resonance (NMR) is used in imaging and spectroscopy techniques throughout chemistry, physics and medicine to determine the structure and concentration of molecules by measuring the magnetic fields created by nuclear spins in the molecules. However, traditional inductive NMR devices have limited sensitivity due to low nuclear spin polarization and large sample-detector distance, and thus require large sample volumes and low spatial resolution to obtain adequate signal. This limitation has motivated the development of new measurement techniques to study NMR at the nanoscale, including microcoil NMR (1), magnetic resonance force microscopy (2), and quantum sensing using solid state electronic spins (3). In particular, the electronic spin of the nitrogen-vacancy (NV) center defect in diamond has been identified as a sensitive nanoscale magnetometer that can operate under ambient conditions and low magnetic fields. In recent work, shallow NV centers at a depth $d$ ranging from nanometers to micrometers under the surface of a bulk diamond chip have been used to measure NMR signals from analytes deposited on the chip (4–15). The NV center's sensitivity to external nuclei improves rapidly for decreasing depth with a dipolar $1/d^6$ scaling, which has enabled single nuclei sensitivity for the shallowest NV centers (16, 17), detection of single proteins (9, 10), and spectral resolution at the level of chemical shifts (12, 13, 18, 19). However, this short-range interaction means that the region of high sensitivity is confined to a small volume directly above the diamond surface (4). This limits high-sensitivity NV-NMR devices in bulk diamond to measuring analytes that can be placed within a few nanometers of the diamond surface.

To overcome the short-range nature of NV-NMR detection, we use recently identified, commercially available, high-purity nanodiamonds with a typical diameter of ~30 nm that host NV centers with stable fluorescence and long coherence time ($T_2 \approx 40$ μs) (20). The small size of these nanodiamond devices allows them to be inserted into relatively thick structures that are inaccessible to other nanoscale NMR techniques, such as inside living cells and organelles (21–24), or other chemical devices with three-dimensional nanostructure such as battery electrodes (25). Nanodiamonds are particularly promising as nanoscale biosensors because they have low toxicity (21), and they support a wide range of surface functional groups which can be engineered for particular applications, such as targeted binding to specific sites in a cell (26). NMR signals from nuclei near small nanodiamonds could provide details about intracellular processes and reveal interactions that are averaged out in conventional ensemble NMR measurements.

In the short term, nanodiamond NV-NMR devices could be applied to complement fluorescent-dye based indicators that are widely used to measure the concentration of chemical species within cells. These dye indicators suffer from several challenges, including bleaching, low chemical specificity, a high toxicity and the need for rigorous calibration from factors like temperature and pH (27, 28). Nanodiamond devices may provide a useful alternative because of their robustness to photobleaching and environmental changes, low toxicity, and their ability to measure multiple species with high specificity.

One limitation of nanodiamond NV-NMR devices is that their geometry varies from device to device: both the nanodiamond shape and the NV center's position within the nanodiamond are uncontrolled in typical fabrication methods. This geometric variability can lead to systematic errors in estimates of analyte properties that are inferred from the NMR signal strength, such as the concentration of the analyte. Here, we address this challenge by *in-situ* self-calibration using the NMR signal created by a layer of nuclei bound to the nanodiamond surface. For example, the few-nanometer thick hydrogen-dense layer observed on oxygen-terminated diamond (6, 7, 29) or a fluorine surface passivation layer (30) could be used for this purpose. We use a Monte Carlo simulation of procedurally generated nanodiamond-NMR device geometries to show that such surface calibration schemes could reduce the systematic error of concentration measurements by 10-fold, making the achievable accuracy comparable to that of the best dye-based techniques.

**Results and Discussion**

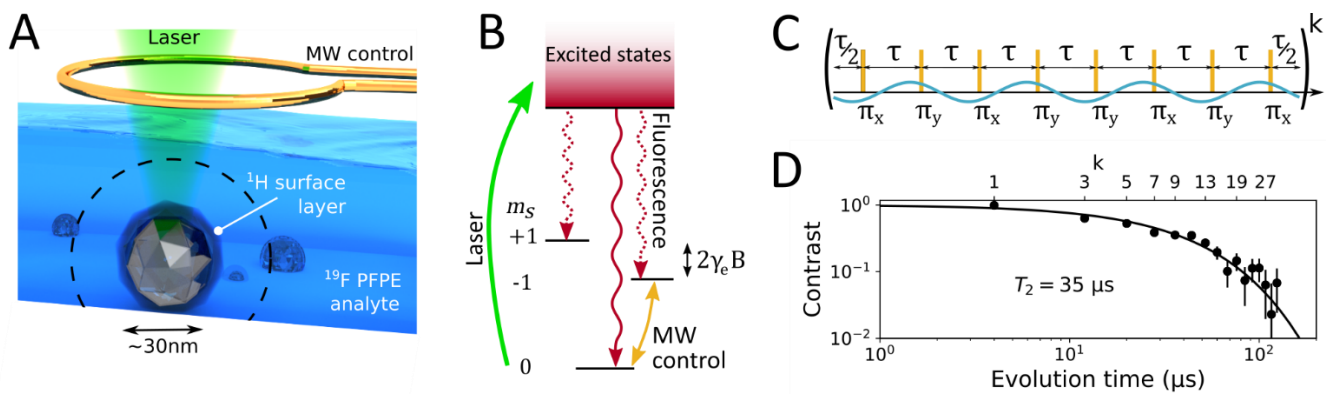

**Fig. 1.** NMR with nanodiamond-hosted NV centers. (A) An illustration of our devices. A nanodiamond (grey) containing an NV center and covered by a hydrogen-dense layer (dark blue) is immersed in the liquid perfluoropolyether (PFPE) analyte environment (light blue). The measured NMR signal is predominantly generated by analyte nuclei close to the nanodiamond, enabling high spatial resolution NMR. Control of the NV center's spin is achieved using a 532 nm laser (green) and a single-loop inductor (yellow) that generates microwave (MW) control pulses. (B) Electronic structure of the NV center. The ground state electronic spin $m_s = \pm 1$ levels are split by $2\gamma_e B$, where $\gamma_e$ is the electronic gyromagnetic ratio and $B$ is the total magnetic field along the NV axis, including both the applied static field $B_0$ and high-frequency components created by local nuclei. The NV electronic spin is initialized by optical pumping induced by the 532 nm laser (green), controlled by microwave frequency pulses (yellow) for spin rotations and readout via fluorescence (red wavy arrows). (C) The CPMG XY8-k dynamical decoupling pulse sequence used to detect NMR signals consists of k repeating XY8 units. The oscillating magnetic field generated by the nuclear spin bath (blue curve) is frequency matched to the decoupling period between microwave $\pi$-pulses (yellow) and leads to a measurable phase of the quantum sensor. The 8-pulse subunit shown is repeated $k$ times to form the complete sequence. (D) A coherence measurement performed on one nanodiamond device using an XY8-k protocol with a constant $\tau = 0.5$ us shows a long $T_2 = 35$us.

To demonstrate the detection of nuclear spins by an NV center hosted in a nanodiamond, we deposited nanodiamonds on a silicon substrate and coated them with a 5 μm-thick layer of perfluoropolyether (PFPE), used here as an example analyte which contains $^{19}$F nuclei. We used decoherence spectroscopy (3, 4) to measure the corresponding NMR spectra. In this method, resonances in the local magnetic field spectrum at the free precession frequency of nearby nuclear spins, $f = \gamma B_0/2\pi$ – where $\gamma$ is the nuclear gyromagnetic ratio, and $B_0$ is the applied static magnetic field – are detected by measuring the decoherence of the NV center's electronic spin during a dynamical decoupling sequence. The measurement protocol proceeds as follows: First, a laser excitation pulse at 532 nm polarizes the NV center into the electronic spin $m_s = 0$ ground state (Fig. 1B). Second, a microwave $\pi/2$-pulse prepares the spin into a superposition state that is sensitive to magnetic fields. Third, an XY8-k sequence (Fig. 1C) decouples the electron spin from magnetic noise while making it sensitive to a narrow spectral band around a center frequency $1/2\tau$ (blue curve in Fig. 1C) with an instrumental bandwidth of $1/8k\tau$. In our experiments, we use a modified form of the XY8-k decoupling sequence (31) to prevent the detection of other spurious frequency bands besides the fundamental at $1/2\tau$ (32). This involves a relative phase shift between the $\pi$-pulses of the XY8 sequence and the $\pi/2$-pulses that constitute initialization and readout of the interferometry sequence (SI appendix). Finally, a second $\pi/2$-pulse maps the electron spin coherence onto the $S_z$ spin basis, which is read out by a second laser pulse. The decoupling period $\tau$ is swept to measure a spectrum of the magnetic field noise power spectral density. The measurement sensitivity of this scheme is determined by the NV-analyte distance and the coherence time of the NV electronic spin, which sets a limit on the useful phase acquisition time. The nanodiamonds used here are ideal for this application because they have a small typical diameter of ~30 nm while still possessing relatively long coherence time (Fig. 1D).

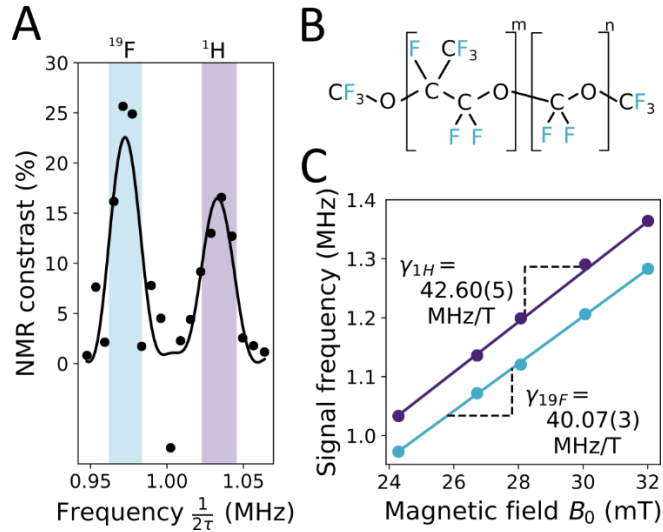

**Fig. 2.** Detection of NMR signals from multiple species with a NV center inside a nanodiamond. (A) An example NMR spectrum measured with an XY8-10 dynamical decoupling sequence. The center positions of the highlight bars correspond to the Larmor precession frequency of $^{19}$F (blue bar) and $^1$H (purple bars). The widths of the bars indicate the filter bandwidth of the XY8-10 protocol. The solid lines show fits to the expected line shape. (B) The molecular structure of the PFPE analyte. (C) The $^{19}$F (blue dots) and $^1$H (purple dots) resonance frequencies extracted from NMR spectra measured as a function of the applied magnetic field $B_0$. Standard error bars are smaller than the data markers. Linear fits are consistent with the gyromagnetic ratios of $^{19}$F and $^1$H.

Figure 2A shows an example NMR spectrum measured with this procedure. The lower frequency peaks (highlighted in blue) are generated by the $^{19}$F nuclear spins in the PFPE analyte at their Larmor frequency. The higher frequency peaks (highlighted in purple) occur at the Larmor frequency of $^1$H nuclei. Pure PFPE itself contains no hydrogen atoms, and we attribute this signal to a few-nm thick hydrogen-dense layer on the surface of the nanodiamond, consistent with measurements on bulk diamond (6, 7, 29). The linewidths of the peaks are consistent with the bandwidth of the decoupling protocol (displayed as the width of the highlight bars), indicating that the intrinsic linewidth of the nuclear spin bath, $1/T_2^*$, is much narrower than the measurement's instrumental linewidth. We extract the signal strength for each nuclear species by fitting the data to the expected lineshapes (15), yielding estimates $\langle B_{1H}^2 \rangle = 0.015(3)$ µT$^2$ and $\langle B_{19F}^2 \rangle = 0.019(1)$ µT$^2$. To confirm that these signals arise from the precession of $^{19}$F and $^1$H nuclear spins, we take several spectra at different applied magnetic fields and extract the resonance frequencies (Fig. 2D). We find that the NMR signal frequencies shift linearly with applied field and have gyromagnetic ratios $\gamma_{1H} = 42.60(5)$ and $\gamma_{19F} = 40.07(3)$ MHz/T, consistent with agreed values.

The decoherence spectroscopy measurement is inherently quantitative. The variance of the magnetic field generated by the nuclear spins $\langle B^2 \rangle$ and the frequency of this field can be determined from the NMR signal, up to the frequency resolution and random errors due to measurement signal-to-noise. However, some quantities that can be inferred from the NMR signal strength – such as the concentration of the analyte or the NV center's distance from the nanodiamond surface – are difficult to measure accurately because the unknown geometry of individual nanodiamonds leads to large systematic errors.

We first illustrate this systematic error by estimating the NV-to-surface distance $d$ and the thickness $t$ of the hydrogen-dense surface layer (shown in Fig. 3A, left) from the measured NMR spectra while taking this geometric variability into account (15). The strengths of the NMR signals provide information about these parameters – stronger overall NMR signals indicate a shallower NV, and stronger $^1$H NMR signals in particular indicate a thicker $^1$H surface layer. To quantify this relationship, we calculate the total magnetic field variance that is measured by an NV center due to an ensemble of nuclear spins (4, 15). This can be expressed as a volume integral of the variance of the dipolar fields created by individual nuclear spins,

$$\langle B_i^2 \rangle = \rho_i \left( \frac{3\mu_0 \mu_i}{4\pi} \right)^2 \int \frac{\cos^2\theta (1 - \cos^2\theta)}{r^6} dV,$$

*(1)*

where the subscript $i$ denotes one of the two relevant nuclear species in our system ($^{19}$F and $^1$H), $r$ is the distance from the NV center, $\theta$ is the polar angle from the NV axis, $\mu_0$ is the vacuum permeability, and $\rho_i$ and $\mu_i$ are the number density and magnetic moment of the nuclei, respectively (4, 15). We assume here that the nuclei have a constant number density within the integration region and that many nuclei contribute to the signal so that the continuum limit is applicable. This result can be rewritten in terms of the NV-to-surface distance $d$ to emphasize the geometric scaling:

$$\langle B_i^2 \rangle = \rho \left(\frac{3\mu_0\mu_i}{4\pi}\right)^2 \frac{S_i}{d^3},$$

*(2)*

where $S_i$ is a dimensionless shape factor that quantifies the effect of the shape of the nanodiamond device on the signal strength. For example, an NV center a depth $d$ below the surface of a bulk diamond chip measuring an analyte on the chip's surface has a small shape factor $S$ because the short range of the dipole-dipole interaction means that only a small volume of the closest nuclei directly above the NV contribute significantly to the signal $\langle B^2 \rangle$. Conversely, an NV center in the middle of a spherical nanodiamond with radius $d$ has a large shape factor because there are equidistant nuclei in all directions contributing to the signal. Both NV centers are the same distance from the diamond surface $d$, but the latter will measure a stronger signal because of the shape of the diamond sensor. The shape factor also includes information about the shape of the nuclear spin region, i.e. the integration region in Eq. (1). For example a very thin $^1$H surface layer compared to the NV-surface distance $d$ has a small shape factor $S_{1H}$. We will now apply Eq. (2) for both $^1$H and $^{19}$F, together with the measured NMR signal strengths and estimated nuclear spin densities, $\rho_{19F} = 40$ spins/nm$^3$ and $\rho_{1H} = 60$ spins/nm$^3$ (6), to infer the geometric parameters $t$ and $d$ for the particular nanodiamond device studied in Fig. 2.

To account for the unknown shape factor in Eq. (2), we perform a Monte Carlo simulation of the nanodiamonds that generates an estimated probability distribution of the shape factors $S_i$. In this Monte Carlo simulation, we procedurally generate 440 different dimensionless nanodiamonds hosting NV centers using a phenomenological algorithm. This algorithm is designed to produce nanodiamond device shapes that have good qualitative resemblance to transmission electron-microscopy images of milled nanodiamonds similar in size to those used here (SI appendix). An example of ten randomly generated nanodiamonds are shown in Fig. 3A. For each generated nanodiamond device shape, we vary the dimensionless ratio of the hydrogen layer thickness $t$ and the NV-surface distance $d$, and calculate the shape factors $S_{19F}(t/d)$ and $S_{1H}(t/d)$ by direct integration. These shape factors, together with the measured values of $\langle B_{1H}^2 \rangle$ and $\langle B_{19F}^2 \rangle$ and Eq. (2) allow us to extract values of $t$ and $d$ that are consistent with both the generated nanodiamond shape and the experimental data.

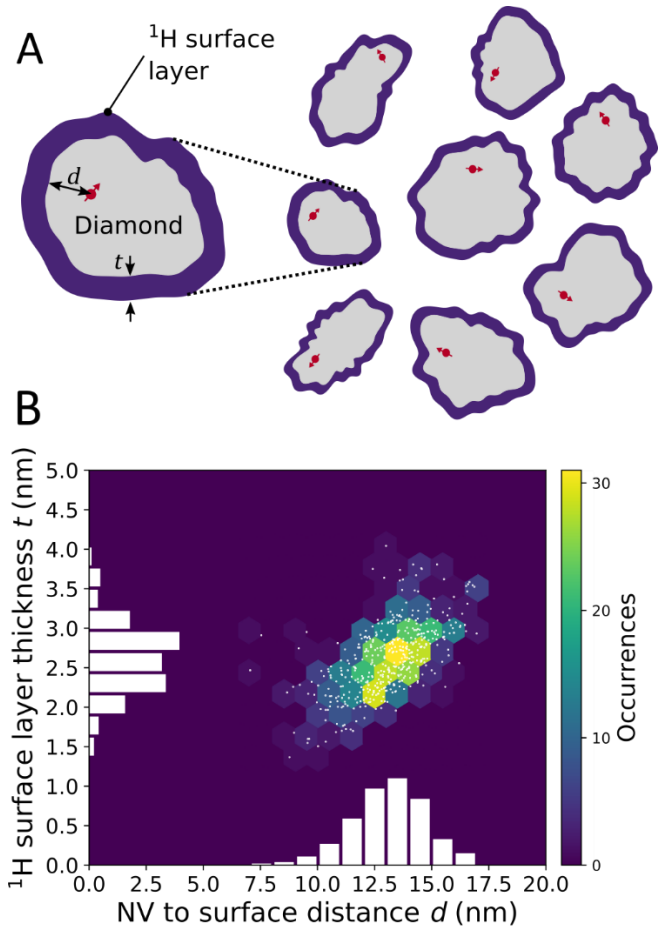

**Fig. 3.** Estimation of the geometry of the measured nanodiamond device using Monte Carlo simulation. (A) Left: A cross section of an example nanodiamond device procedurally generated by the Monte Carlo simulation, illustrating the geometric parameters $d$ (the distance between the NV center and the nanodiamond surface) and $t$ (the thickness of the hydrogen surface layer). We generate the nanodiamond shapes using a realistic model for surface roughness and place an NV center within the each nanodiamond (SI appendix). Right: eight examples of generated nanodiamond devices out of the 440 instances used for simulation. (B) Histogram of the values of the geometric parameters extracted from the experimental data and Monte Carlo simulation. Each individual nanodiamond device shape generated by the simulation is plotted as a white dot, with coordinates indicating the values of the $t$ and $d$ parameters that are consistent with both the measured data and the shape of that generated nanodiamond device. The single-parameter histogram for the surface layer thickness $t$ (NV to surface distance $d$) is plotted as white bars on the vertical (horizontal) axis.

Figure 3A displays the results of this simulation as a two-dimensional histogram of the extracted values $t$ and $d$. Each white dot plots the values of $t$ and $d$ extracted for one of the generated nanodiamond device shapes. The distribution created by these values represents our best estimate for the probability distribution of the values of $t$ and $d$ for the nanodiamond measured above while including the systematic error due the unknown shape of the nanodiamond. We note that this distribution is robust to moderate changes in the algorithm we use to generate the nanodiamond shapes (SI appendix).

The histogram of extracted geometric parameters in Fig. 3B shows that this device has a small NV center depth $d \approx 13.2 \pm 1.6$ nm, where here the error is the standard deviation due to the geometric variability. A second device we used to measure NMR signals has $d \approx 11.4 \pm 1.3$ nm (SI appendix, Fig. S3), confirming that these devices provide the high spatial resolution of NV centers close to the nanodiamond surface. We note that the systematic error caused by the unknown nanodiamond shape – the standard deviation of $d$ in the histogram in Fig. 3B – is only about fourfold larger than the error from the measurement signal-to-noise ratio.

Having established the nanoscale size and measurement capabilities of these devices, we now describe how the NV-NMR measurement data can be used to probe the local nuclear spin concentration of an unknown analyte. The sensitivity to external analyte spins $m = \langle B^2 \rangle / \rho$ can differ from device to device by more than 10-fold because of geometric variability, making it difficult to infer the nuclear spin concentration from the NMR signal measured by a single device.

There are multiple approaches offering possible solutions to this calibration challenge. One direct approach is to measure each device's sensitivity $m$ individually by performing an NMR measurement on a solution of known concentration with the device before transferring that device to the unknown analyte. However, methods for tracking and manipulating individual nanodiamonds are cumbersome, which makes such a pre-calibration approach challenging and demands the capability for *in-situ* calibration. In a second approach we call batch calibration, a subset of the devices in a batch could be used to measure the average sensitivity $\langle m \rangle$ using a known solution. Another device from the batch would then be used to measure the unknown analyte, and the analyte spin density could be estimated using the batch average $\langle m \rangle$ as the sensitivity. While this approach avoids the challenge of manipulating single nanodiamonds, the estimated spin density would be inaccurate because of the large deviations of each device's sensitivity $m$ from the batch average $\langle m \rangle$.

We propose instead a third approach that we call surface-enhanced self-calibration. The idea is to engineer nanodiamond devices with a thin coating of nuclear spins, much like the naturally occurring hydrogen-dense layer, and use the NMR signal from these surface spins to help estimate the sensitivity $m$ to external spins of each individual device *in-situ*. The NMR signal strength of the surface spins alone does not completely determine the sensitivity $m$, but it provides substantial information that can be used to improve the estimates of unknown analyte densities. This surface-enhanced calibration scheme requires that the termination species' NMR signal be spectrally distinguishable from analyte NMR signals, and that the concentration and thickness of the termination layer be relatively consistent. Well-developed techniques for nanodiamond surface termination with a variety of nuclear species suggest these requirements can be met feasibly (26).

We first examine the surface-enhanced calibration scheme analytically in the limit of a thin surface layer. In this limit, the surface layer shape factor and the analyte shape factor can be approximated as $S_s \approx S'_s \frac{t}{d}$ and $S_a \approx S_a\left(\frac{t}{d} = 0\right)$, respectively, where $t$ is now the thickness of the surface layer, and $S'_s$ is the dimensionless slope of $S_S\left(\frac{t}{d}\right)$ for thin layers. The density of the unknown analyte can be estimated from the measured NMR signal strengths of the analyte nuclear spins with unknown density and the surface spins by applying Eq. (2) for both the surface and analyte spins, yielding

$$\rho_a = C\ \left(\frac{4\pi}{3\mu_0}\right)^{\frac{1}{2}} \frac{\langle B_a^2 \rangle}{\mu_a^2} \left(\frac{\mu_s^2}{\langle B_s^2 \rangle}\right)^{\frac{3}{4}},$$

*(3)*

where the subscripts $a$ and $s$ refer to the surface and analyte nuclear spins and $C = (\rho_s t S'_s)^{3/4}/S_a$ is the nanodiamond-specific calibration factor, which depends on the surface spin density $\rho_s t$ and the dimensionless shape parameters. The calibration factor $C$ cannot be determined for a particular device in a measurement on an unknown analyte, so, as in the batch calibration scheme, we can instead use the batch average calibration parameter $\langle C \rangle$ to produce an estimate for the analyte density $\rho_a$. Deviations of $C$ from the ensemble average $\langle C \rangle$ will introduce error into the estimate for $\rho_a$. However, this error will generally be much smaller than the error in the batch-only calibration scheme that is based on only the analyte signal. This is because the calibration parameter for the batch calibration scheme, namely the sensitivity $m$, has a strong cubic dependence on the NV-to-surface distance $d$, as seen in Eq. (2). This $d^3$-dependence is eliminated from the calibration parameter in the surface enhanced calibration. Furthermore, the shape factor term $S'^{\frac{3}{4}}_s/S_a$ in the surface-enhanced calibration factor $C$ contributes a relatively small error because although $S'_s$ and $S_a$ can vary by more than 10-fold, these shape parameters are also strongly correlated for typical nanodiamonds (SI appendix).

To quantify the effectiveness of the surface-enhanced calibration scheme, we perform Monte Carlo simulations of NV-NMR measurements on unknown analytes. In these simulations, nanodiamond devices are procedurally generated as described earlier, but with a dimensional size pulled from the measured probability distribution of nanodiamond diameters for our devices (20). We simulate NMR measurements by calculating the signal strengths $\langle B_a^2 \rangle$ and $\langle B_s^2 \rangle$ using Eq. (1) for each of the 440 generated nanodiamond devices.

Figure 4A shows the results of a simulation in the ideal situation where the calibration layer thickness is the same $t = 1$ nm for every nanodiamond. The estimated analyte density $\rho_a$ follows a log-normal distribution with a relative standard deviation of 224% when using the batch calibration scheme. In contrast, our surface-enhanced calibration reduces the relative standard deviation to only 24%. The accuracy after surface-enhanced calibration is comparable to that which can be obtained with ratiometric dye indicator techniques for concentration sensing, without the need for extensive pre-calibration against environmental parameters like pH and temperature (28, 33, 34). To confirm that the surface-enhanced self-calibration scheme is robust to small variability in the surface layer thickness, we repeat the simulation, but allow the thickness of the surface layer to vary between generated nanodiamonds in a log-normal distribution. Figure 4B shows that even if the surface layer thickness has a large relative standard deviation of 50%, the surface-enhanced calibration scheme improves accuracy by 4-fold. These results suggest that surface-enhanced calibration could be a useful scheme to accurately measure the local concentration of nuclear spins with nanoscale resolution.

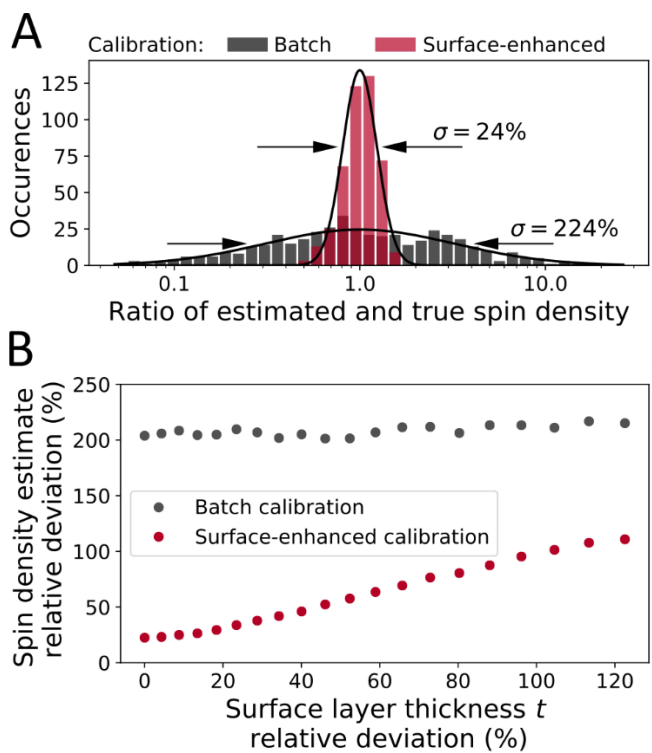

**Fig. 4.** Comparison of different calibration methods for the nanodiamond devices. (A) Histogram of simulated analyte spin density measurements for different nanodiamond geometries, with the surface layer thickness $t$ held constant at $t = 1$ nm. Batch calibration shows large systematic errors due to uncontrolled geometry variation. The surface-enhanced calibration scheme reduces systematic error by almost a factor of 10. (B) We perform simulations like those shown in (A), but allow the surface layer thickness $t$ to vary between nanodiamonds in a log normal distribution with geometric mean of $\bar{t} = 1$ nm. When $t$ has a large standard deviation, this source of noise reduces the measurement accuracy.

In conclusion, our results demonstrate that NV centers hosted inside small nanodiamonds can be used to detect NMR signals from multiple nuclear species in a sample volume of $(19 \pm 4 \text{ nm})^3$ (SI appendix). Further, the surface-enhanced calibration technique we propose here largely corrects the systematic errors caused by geometric variability of nanodiamond devices and enables the decoding of the measured NMR signal. With this added capability, such devices could be applied to complement fluorescent-dye based indicators that are widely used to measure the concentration of chemical species within cells. As NV-NMR techniques continue to improve, it may become possible to perform single molecule structural determination *in vivo* (35–37). We anticipate that nanodiamond NV-NMR devices, combined with further efforts to improve NV coherence in small nanodiamonds (20) and the use of enhanced readout techniques (38–40), will enable powerful new methods for the study of cellular structures and processes.

**Materials and Methods**

We use a scanning confocal microscope (NA = 0.9 air objective lens) to isolate individual nanodiamond-hosted NV centers. A 532-nm laser was modulated by an AOM (Acousto-Optical Modulator, AA optoelectronics MT80-A1-VIS) to create pulses. The laser intensity was set to the saturation power of the NV center, and 850-ns and 350-ns pulses were used to initialize and readout the electronic spin state, respectively. NV center fluorescence during readout was filtered to select the 600 to 800 nm band and coupled into a single-mode optical fiber and detected by an avalanche photodiode (Excelitas SPCM-AQRH-14-FC) that was time-gated to select counts during readout by a switch (MiniCircuits ZASWA-2-50DR+). The static magnetic field was created by a neodymium permanent magnet mounted on a two-axis rotation mount, centered on the sample.

Microwave pulses were generated by a Tektronix 70002A 10 GHz Arbitrary Waveform Generator, then amplified (Mini-Circuits ZHL-16W-43+) and terminated in a ~2-mm diameter single-loop inductor mounted between the sample and objective.

The nanodiamonds were purchased from Nabond Technologies Co. and dispersed in a solution of 10 mg of nanodiamond powder to 25 mL of ethanol via sonication. This solution was deposited on a quartz substrate using an Omron U22 nebulizer. This method produces a well dispersed distribution of nanodiamond locations and allows individual addressing by the confocal microscope. We spin-coated the quartz substrate and nanodiamonds in a 5 μm thick layer of Fomblin Y brand perfluoropolyether (PFPE) oil with average molecular weight of 6600 amu.

**Acknowledgements**

We thank Gavin Morley and Benjamin J. Woodhams for stimulating discussions. We gratefully acknowledge financial support by the Leverhulme Trust Research Project Grant 2013-337, the European Research Council ERC Consolidator Grant Agreement No. 617985, the EPSRC Quantum Technology Hub NQIT (EP/M013243/1) and the Winton Programme for the Physics of Sustainability. H.S.K. acknowledges financial support by St John's College through a Research Fellowship. J.H. acknowledges the UK Marshall Aid Commemoration Commission for financial support.

# Supplementary Information for

Nanoscale NMR Spectroscopy using Self-Calibrating Nanodiamond Quantum Sensors

Jeffrey Holzgrafe[1,2], Qiushi Gu[1], Jan Beitner[1], Dhiren Kara[1], Helena S. Knowles[1,3], and Mete Atatüre[1†]

[1] *Cavendish Laboratory, University of Cambridge, JJ Thomson Avenue, Cambridge CB3 0HE, UK*
[2] *John A. Paulson School of Engineering and Applied Sciences, Harvard University 29 Oxford Street, Cambridge MA 02138, USA*
[3] *Department of Physics, Harvard University, 17 Oxford Street, Cambridge MA 02138, USA*

Corresponding author: Mete Atatüre
Email:  ma424@cam.ac.uk

**This PDF file includes:**

>   Supplementary text
>   Figs. S1 to S6
>   References for SI reference citations

**Supplementary Information Text**

**Discriminating 13C spurious harmonics with a modified XY8-k sequence.** Detection and identification of NMR signals is complicated by the spurious harmonics of the CPMG XY8 sequence's filter function (1, 2). In particular, the spurious $4f$-harmonic of the $^{13}$C Larmor frequency (where $f$ is the frequency of the magnetic field signal) is separated from the $^1$H Larmor frequency by only 0.6%. This strong overlap makes it difficult to ascribe a signal detected near this frequency to external $^1$H, rather than the $^{13}$C present within the diamond lattice (natural abundance: 1.1%). These spurious harmonics are caused by the finite length of control pulses in the implementation of the XY8 protocol. A magnetic field signal changes the electron spin resonance frequency, which changes the Bloch rotation axis of a control pulse at time $t_{\text{pulse}}$ by an angle $\phi_{\text{spur.}} \sim B_z(t_{\text{pulse}})\gamma_e/\Omega_R$, where $B_z$ is the time-dependent part of the magnetic field along the axis of the NV center and $\gamma_e$ is the electron spin gyromagnetic ratio, and $\Omega_R$ is the Rabi frequency of the control pulses. If the magnetic field signal is frequency-matched with the intentional changes in the phase of the decoupling $\pi$-pulses that are used as part of the CPMG XY8 sequence, then the small spurious rotations of the state by $\mathcal{O}(\phi_{\text{spur.}})$ can add constructively, contributing an effective spurious phase shift to the measurement (1).

Here, we use a recently proposed modification to the XY8 protocol which allows us to decouple from the 4f spurious harmonic (2). This modified XY8 protocol uses a different Bloch sphere rotation axis for the initial preparation of a superposition state than the rotation axis for the decoupling $\pi$-pulses. By properly tuning the preparation Bloch sphere rotation axis, the protocol can become less sensitive to spurious harmonics, while having a negligible effect on the conventional harmonics (2). It was shown that the NMR contrast of the 4f spurious signal, $C_{spur}$ depends on the angle $\nu$ between the preparation and decoupling Bloch axes, as

$$C_{spur} \propto (\sin(2\nu) - 1)\varphi_{spur}^2 + \mathcal{O}(\varphi_{spur}^3).$$

We verified this prediction by performing an NV-NMR measurement on a nanodiamond device without any explicit analyte coating (i.e. just the naturally occurring ¹H). In this measurement, we hold the pulse spacing $\tau$ fixed to be on-resonance with the $1f$-harmonic of ¹H and the $4f$-harminc of ¹³C NMR signals and sweep the preparation angle offset $\nu$. The results, shown in Fig. S1A, fit well to the expected sinusoid dependence. Fig. S1B shows that the full NMR signal spectral peak in the spurious-signal maximizing regime ($\nu = 45°$) is both taller – indicating a larger total phase acquisition – and wider – indicating a spectrally broad source of noise – than the spectrum observed in spurious signal minimizing regime ($\nu = 135°$). These results suggest that our experiments occur in a regime where both the ¹³C spin bath in the diamond host, which is broadened by hyperfine interaction with the NV center's electronic spin, and a ¹H spin bath outside the nanodiamond can be detected, depending on the preparation angle offset $\nu$ of the modified XY8-k sequence. To clarify the detection of external nuclear spins, all the experiments described in the main text are performed in the spurious-minimizing regime with $\nu = 135°$.

We confirm that the spurious harmonic 13C signal does not contribute to the 1H signal when the protocol is set to its minimum sensitivity to spurious harmonics ($\nu = 135°$) in two ways. First, we verified that the gyromagnetic ratio of the NMR signal is statistically consistent with 1H and inconsistent with 13C (Fig. 2C of main text). Second, we performed an experiment to estimate the contribution to the signal from the $4f$-harmonic. In this experiment, we use an artificial noise signal that can be picked up by the $4f$-harmonic to calibrate the $4f$-harmonic response. This artificial noise signal is generated by sending a 400 kHz current through the same inductive loop that generates our control pulses. We quantify the amplitude of the artificial noise $B_a$ along the NV center's z-axis in units of magnetic field by performing an NV-NMR measurement tuned to detect 400 kHz signals as a $1f$-harmonic and applying the correct relationship between the NMR signal contrast and the magnetic field amplitude for a noise signal with constant amplitude and random phase (3). Now tuning the NV-NMR measurement to detect the 400kHz signal as a $4f$-harminc and sweeping the artificial signal amplitude while monitoring the NMR contrast, we find the expected $C_{spur} = kB_a^3$ dependence (shown in Fig. S2, bottom), where $k$ is a proportionality constant. This proportionality constant $k$ is independent of the signal frequency in the limit $2\pi f \ll \Omega_R$ because the spurious phase acquired during the XY8 sequence depends only on the magnetic field at the time of the control pulses. Thus, we can conclude this dependence will hold for the ¹³C spurious signal as well, with the same numeric value of $k$. We use this spurious signal calibration to estimate the size of the spurious signal in the 1H NMR measurements shown in the main text. The NMR contrast expected from a quasistatic $^{13}$C bath is

$$C_{spur} = \int P(B_a)\, k\, B_a^3\, dB_a,$$

where $P(B_a)$ is the probability distribution for the quasistatic magnetic field signal amplitude generated by the $^{13}$C bath. We assume $P(B_a)$ to be a gaussian distribution and measure the variance from a spin-echo measurement tuned to the $^{13}$C Larmor frequency (4). The resulting distribution $P(B_a)$ is shown in Fig. S2, top. Performing the above integral, we estimate that less than 1% of the observed NMR signal strength is due to the spurious $^{13}$C dip.

**NMR signal extraction.** All pulse protocols described here were performed simultaneously with a reference pulse protocol wherein an extra microwave $\pi$-pulse is used before readout. The difference in fluorescence between the standard and reference pulse creates a robust raw dataset. We fit measured NMR signal data to the expected lineshape described by Pham et. al. (5), using the long nuclear-spin bath coherence limit when appropriate. The full fit function is the product of two lineshapes for the hydrogen and fluorine spin baths and a stretched exponential envelope function used to model the effects of other noise sources on the spin coherence. The plotted data is normalized to this envelope function to produce the NMR contrast shown in the figures.

**Monte Carlo simulations.** As described in the main text, the geometric variability of our nanodiamond devices can add substantial systematic error to parameter values inferred from measurement data. To quantify the magnitude of this systematic error, we employ a Monte Carlo simulation method which uses an ensemble of many procedurally generated nanodiamond geometries to estimate a distribution of parameter values. The main text describes the results of two different Monte-Carlo simulations. We use the first simulation to determine the values of the NV to surface distance $d$ and hydrogen layer thickness $t$ from the experimental data and the dimensionless nanodiamond shape for each generated nanodiamond in the ensemble. In the second simulation, we generate full dimensional nanodiamond devices and simulate NV-NMR measurements using these devices. We then use the results of these simulated measurements to compare the calibration methods described in the main text. We will now describe the procedure for each of these simulations in detail.

The first simulation, for determining the values of the geometric parameters $t$ and $d$, proceeds in the following steps. See also Fig. S3.
1. We generate a dimensionless nanodiamond surface shape. The outer surface of the nanodiamond is first generated by starting with a random ellipsoid with aspect ratios constrained between 1:1 and 1:3 and adding surface roughness. Our surface roughness model uses a sum of spherically smooth periodic functions with gaussian amplitude distribution and random phase to generate a surface roughness with rms amplitude of roughly 15% of the nanodiamond diameter, and with a power spectrum that falls off as the inverse of the number of periods per circular revolution. The mathematical form of the surface roughness model was chosen to qualitatively reproduce the nanodiamond shapes observed in electron microscopy imaging of high-pressure, high-temperature (HPHT) grown nanodiamonds of similar size to the ones used in this work (6, 7). To verify that our simulation results are robust, we repeated the full simulation with different choices of nanodiamond generation procedures. We found that simple surface geometries such as smooth spheres and ellipsoids produce joint distributions of $t$ and $d$ that display noticeably stronger correlation than our chosen surface geometry, which motivated the inclusion of roughness into the simulation. For reasonable models that include surface roughness, we found that the choice of model had only a minor impact on the resulting distributions.

2. We procedurally place an NV center within the nanodiamond such that all locations are equally likely except those within 15% of the mean nanodiamond diameter of the surface (a distance of roughly 5nm for nanodiamonds of the size studied in this work) to account for the fact that near surface NVs have poor charge state stability and coherence. The orientation of the NV center is chosen such that all orientations are equally likely.
3. We calculate the shape factors of this NV-nanodiamond system for the $^{19}$F analyte and $^1$H surface regions, as described in the main text. We distinguish between three regions around the nanodiamond: the diamond region extends from the origin to the diamond surface generated earlier, the 1H surface region extends a distance $t$ outward from the diamond surface (for simplicity, we take the outer surface of the hydrogen layer to extend past the diamond surface a distance $t$ directed from the origin), and the $^{19}$F analyte region, which we assume extends out infinitely far. Because the thickness of the $^1$H surface region is unknown, we calculate the shape factors for several values of the dimensionless thickness parameter $t/d$ and interpolate to determine the shape factors as a function of the surface thickness parameter: $S_{19F}(t/d)$ and $S_{1H}(t/d)$.
4. We use the experimentally measured magnetic field variances and simulated shape factors to solve the system of equations made up of Equation 2 in the main text applied to both the $^{19}$F analyte and the $^1$H surface spins for the geometric parameters $t$ and $d$.
5. Steps 1-4 above are repeated for many different hypothetical nanodiamonds. The distribution of the calculated values of the geometric parameters $t$ and $d$ (Fig. 3B of main text) provides an estimate for the joint probability distribution of $t$ and $d$, while taking into account the systematic error introduced by geometric variability.

The second simulation, to compare calibration methods, proceeds similarly:
1. We procedurally generate a hypothetical nanodiamond devices similar to steps 1-2 in the first simulation but include dimensional information in the geometry of the device. The geometric mean diameter of the hypothetical nanodiamond is generated by a gaussian distribution with a mean of 23 nm and a standard deviation of 7nm, which matches the distribution of nanodiamond diameter for the nanodiamond ensemble used in these measurements that was previously measured by AFM (8). Furthermore, we generate the surface layer thickness $t$ either by fixing it to exactly 1nm (Fig. 4A of the main text) or by generating it from a log-normal distribution with a geometric mean of 1 nm and a variable deviation (Fig. 4B of the main text).
2. We calculate the magnetic field variance that would be measured for this hypothetical nanodiamond device, as described in the main text.
3. We repeat steps 1-2 above for many hypothetical nanodiamonds. For Fig. 4B of the main text, each generated nanodiamond geometry is resampled ten times with a new layer thickness to precisely estimate the impact of a variable surface thickness. These results simulate an experiment in which many nanodiamonds with an engineered surface layer are used to measure NMR signals of a known analyte.
4. We calculate estimates of the analyte density based on the simulated measurement, using both the naïve and the surface-enhanced calibration methods described in the main text. We determine the best average calibration factors for the naïve and surface-enhanced calibration methods ($\bar{C}'$ and $\bar{C}$, respectively) by ensuring that the geometric mean of the estimated analyte concentration for all the generated nanodiamonds is equal to the true analyte density. This type of overall ensemble calibration could be performed in practice by measuring a known analyte using a representative sample of the nanodiamond ensemble. The distribution of the analyte density estimated by the hypothetical nanodiamond devices tells us how much systematic error the geometric variability adds when using the different calibration methods.

**Correlations between surface and analyte shape parameters.** To investigate the correlations between the shape parameters for typical nanodiamonds, we used the generated nanodiamond shapes described above to calculate the shape parameters $S_a$ and $S'_s$ (see main text for definition). The results shown in Fig. S4 indicate that these parameters are highly correlated, with a coefficient of determination $r^2 = 0.92$. This strong correlation improves the surface enhanced calibration scheme by reducing the deviations of the calibration parameter $C$ from the ensemble average due to the nanodiamond shape.

**Sample volume estimation.** We estimate the sample volume using the Monte-Carlo integration method described by Staudacher et. al. (9). Briefly, in this method we randomly place $N$ points representing analyte spins in a large volume outside the nanodiamond (with dimensions $t$ and $d$ determined as described above), and calculate the contribution of each point to the magnetic field variance sensed by the NMR measurement. The sample volume is estimated by

$$V_{\text{sample}} = \frac{N_{50}}{N} V,$$

where $N_{50}$ is the minimum number of points required to generate 50% of the total magnetic field variance, and $V$ is the total volume in which the points are placed. We perform this procedure on a subset of the nanodiamonds generated in the first Monte-Carlo simulation described above, which was used to estimate the geometric parameters $t$ and $d$. A histogram of the calculated sample volumes (Fig. S3) allows us to estimate a probability distribution for the sample volume while including the geometric variability. From this data we estimate the sample volume to be $(19 \pm 4 \text{ nm})^3$, where the error represents the standard deviation limited by geometric variability.

Using this sample volume calculation, we further estimate that the signal is produced by of order $40 \ \frac{\text{spins}}{\text{nm}^3} * 19^3 \text{ nm}^3 = 3 * 10^5$ spins of $^{19}$F, contained in about 1500 PFPE molecules of 6600 amu.

**Measurements on other nanodiamonds.** We performed NV-NMR measurements on two other nanodiamond devices besides the one described in the main text. One of these was not covered in an explicit analyte; some of the measurement results are shown in Fig. S6A. The final nanodiamonds device was prepared identically to the one described in the main text. Figure 3 shows an example NV-NMR spectrum produced by measurements on this device. The $^1$H NMR signal for this device is substantially broader than the limit provided by the XY8-10 protocol, which is denoted by the width of the blue highlight bar in the figure. This indicates that the $^1$H nuclear spin bath has a relatively large dephasing rate. The hydrogen layer thickness inferred from the NMR signal strength (Fig. S6B) using the method described in the main text is also substantially thicker than that of the device described in the main text.

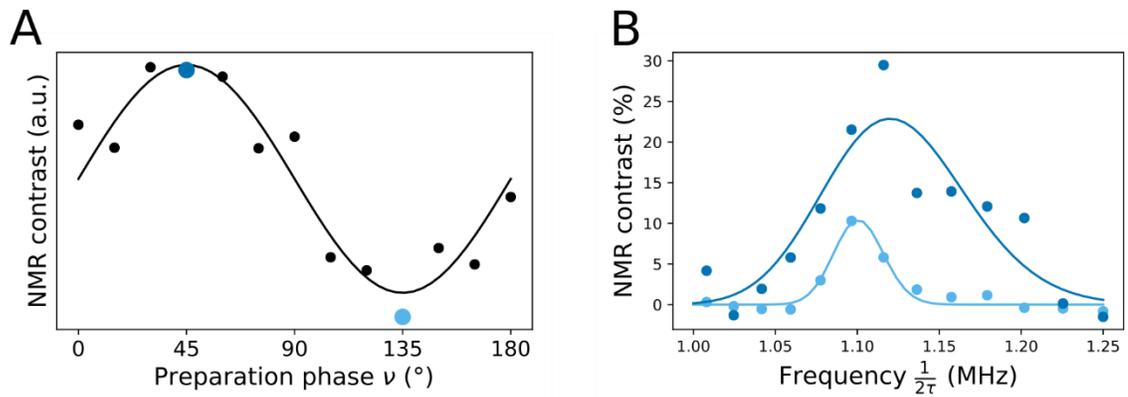

**Fig. S1.** Control of the spurious signal sensitivity with preparation angle for an XY8-8 protocol. (A) The NV coherence after an XY8-8 sequence with 16ns long pi pulses on-resonance with the $1f$-harmonic $^1$H or $4f$-harmonic $^{13}$C- NMR signals. The data matches with the expected dependence (black line). Light (dark) blue points show the NMR spectra when the protocol is tuned to be minimally (maximally) sensitive to the $^{13}$C-4f signal. (B) NV-NMR spectra of the maximally and minimally 4f-sensitive protocols. Other parameters identical to those in A).

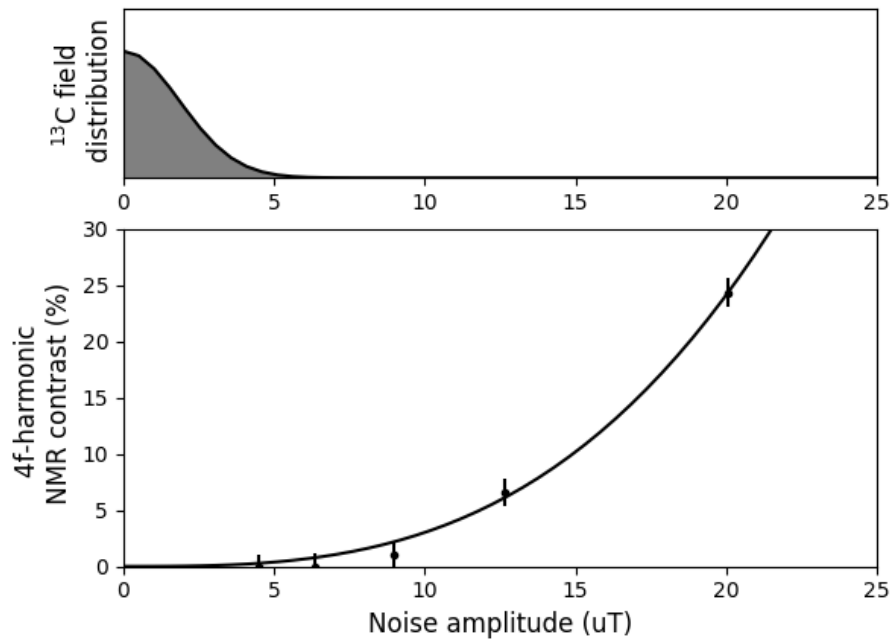

**Fig. S2.** Calibration of the spurious $4f$-harmonic NMR signal strength of an XY8-8 measurement under the minimum sensitivity $v=135°$ condition. Bottom: The spurious $4f$-harmonic NMR contrast generated by an artificial magnetic field signal with amplitude $B_{a0}$ (black points) fits to the expected $B_{a0}^3$ dependence (black line). Top: Probability distribution of the quasistatic magnetic field signal amplitude generated by the $^{13}$C nuclear spin bath, inferred from measured spin-echo signal.

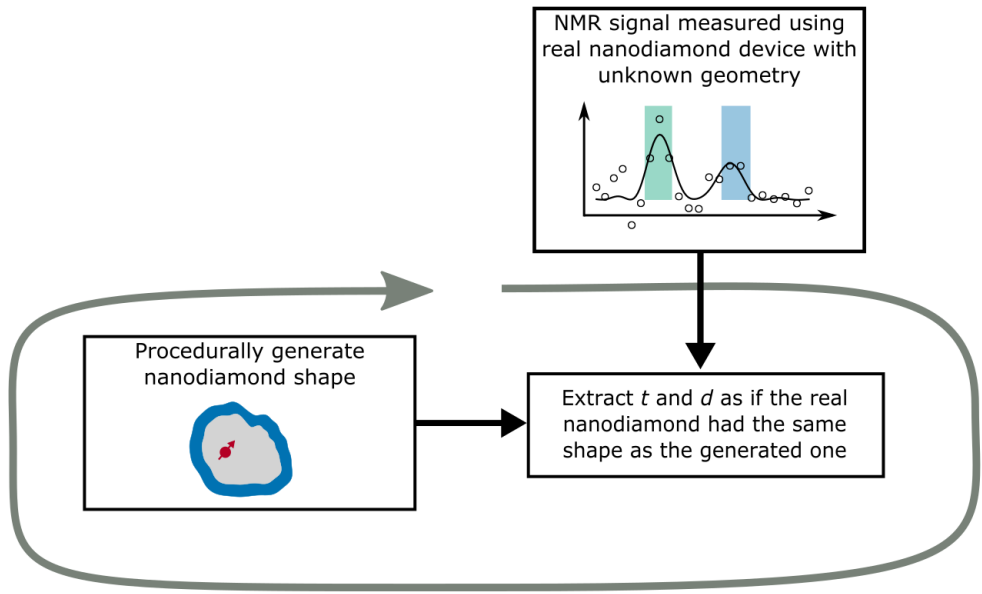

**Fig. S3.** Schematic of the procedure to extract the geometric parameters for the measured nanodiamonds.

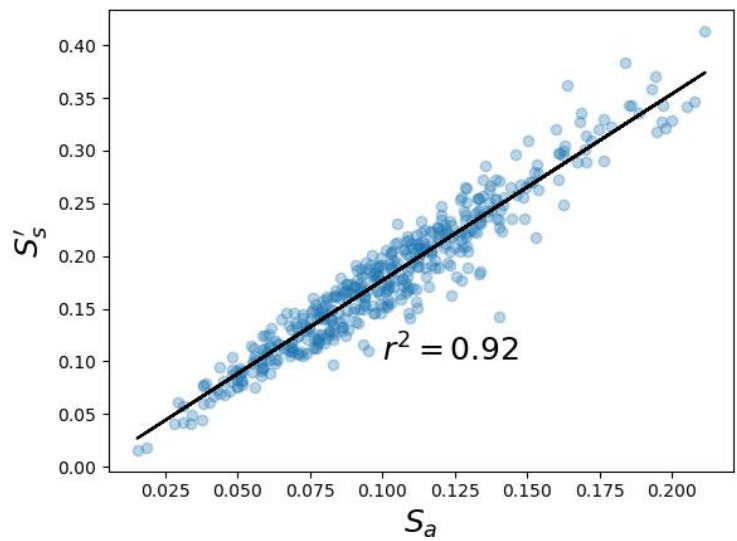

**Fig. S4.** The calculated surface layer and analyte shape parameters ($S'_s$ and $S_a$ – see main text) for different simulated nanodiamond devices (plotted as blue dots) shows a strong linear correlation (black line).

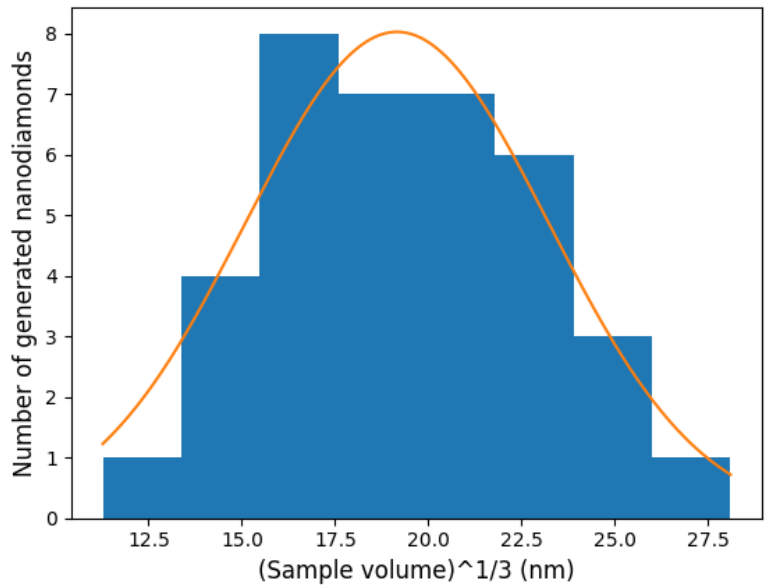

**Fig. S5.** A histogram (blue) of the sample volumes inferred from the experimental data and the simulation of generated nanodiamonds. The orange line shows a gaussian fit with mean of 19 nm and a standard deviation of 4 nm.

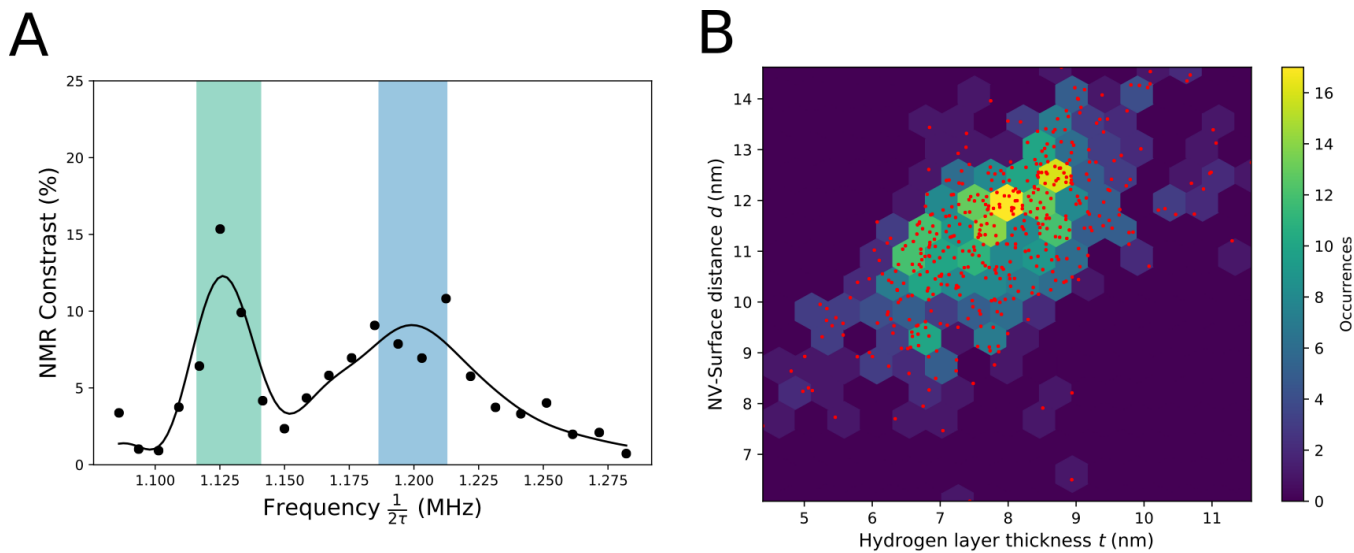

**Fig. S6.** NV-NMR measurements on another nanodiamond device. A) NMR spectrum measured with an XY8-10 dynamical decoupling sequence at $28.2\ mT$. The positions of the highlight bars correspond to the Larmor precession frequency of 19F (green bars) and 1H (blue bars). The width of the bars indicates the filter bandwidth of the XY8-10 protocol. The NMR spectrum of this nanodiamond device shows a much broader hydrogen signal spectral width, indicating a faster 1H nuclear spin bath dephasing time. B) The distribution of the geometric parameters $t$ and $d$ inferred using the method described in the main text.